# ON THE WAVE ENERGY POTENTIAL OF THE BULGARIAN BLACK SEA COAST

**Vasko Galabov**[1]

[1] National Institute of Meteorology and Hydrology- Bulgarian academy of sciences, **Bulgaria**
(For correspondence: vasko.galabov@meteo.bg )

**ABSTRACT**

In the present study we evaluate the approaches to estimate the wave energy potential of the western Black Sea shelf with numerical models. For the purpose of our evaluation and due to the lack of long time series of measurements in the selected area of the Black Sea, we compare the modeled mean wave power flux output from the SWAN wave model with the only available long term measurements from the buoy of Gelendzhik for the period 1997-2003 (with gaps). The forcing meteorological data for the numerical wave models for the selected years is extracted from the ERA Interim reanalysis of ECMWF (European Centre for Medium range Forecasts). For the year 2003 we also compare the estimated wave power with the modeled by SWAN, using ALADIN regional atmospheric model winds. We try to identify the shortcomings and limitations of the numerical modeling approach to the evaluation of the wave energy potential in Black Sea.

**INTRODUCTION**

During the last years there is an increasing research on the possibilities to extract energy from the oceans and that includes mainly the wave energy conversion, which is now already considered a commercial technology [1-3] especially in some high energetic areas of the Atlantic Ocean. One of the main tasks with regards to the wave energy utilization is the preliminary step- estimation of the available energy potential. Extensive studies have been performed especially for some areas that are known as the most energetic in Europe, like Portugal [4-7] and Spain [8-11]. Our opinion is that the mentioned here studies can be taken as an examples of a good practices on the problem of wave energy evaluation. These studies typically combine the numerical modeling with studies of the available extensive and long term in-situ measurements of the wave parameters that helps to produce reliable estimations.

For the Black Sea contrary to the mentioned study sites, there is an almost total lack of measurements of the wave parameters with some exceptions- some wave measurements performed in the frame of NATO TU-WAVES project. The only long enough measurements are for the buoy near to Gelendzhik, Russia (for the period 1996-2003 with some gaps- available at www.coastdyn.ru, provided by the Russian oceanographic institute Shirshov). There are also some other measurements at the Gloria oil platform in the Romanian shelf and some for the Galata gas platform nearby Varna, Bulgaria (but we don't have an access to these measurements and therefore we can't use them in our present study). In such situation not surprisingly the only option in order to evaluate the wave energy potential is the use of numerical wave models, while the measurements have been used for model validations. The first study on the wave energy estimations in the Black Sea was published by Rusu [12]. In this article Rusu presented an estimation of the wave climate, based on measurements in the Romanian shelf, validated SWAN wave model [13] and estimated the wave energy patterns for some moderate and high energetic conditions. Two recent articles- by Akpinar and Komurcu [14] and Aydogan et al [15] provide estimations of the Black Sea wave energy potential. The study of Akpinar and Komurcu is based on the usage of SWAN model, forced by ERA Interim [28] data (with a spatial resolution of 0.75) for a period of 15 years. The study of Aydogan is based on the use of MIKE21 model (a commercial model developed by DHI company) based on unstructured grid approach and forced according to the authors by ECMWF winds with a spatial resolution of 0.1° for 13 years (it is not mentioned in the article what kind of model output- downscaling of reanalysis or of the operational product). The two studies estimations are very different- while the study of Akpinar and Komurcu concludes that the maximal mean wave power potential of the Black Sea is up to 3kW/m, according to Aydogan it is 7kW/m and also the wave power potentials of the points at the Western Black Sea shelf differ significantly- from 1.5-2 kW/m according to Akpinar to 4.8 according to Aydogan. The present study try to answer the question about the source of this significant discrepancy and to present our preliminary evaluation of the usability of different wind sources for a detailed wave power potentials evaluation for the Bulgarian shelf area.

**METHODOLOGY**

The present study is based on the usage of the SWAN wave model. The SWAN implementation is actually the operational version of SWAN of the National Institute of Meteorology and Hydrology of the Bulgarian Academy of Sciences (NIMH-BAS). Details about our implementation of SWAN can be found in [16]. The spatial resolution of SWAN for that study is 2 arc-minutes and the output of the wave parameters is for every 1 hour. The wave generation/dissipation parameterization is the parameterization formulation proposed by Westhuysen et al [17] based on the wind input of Yan [18] and the saturation based whitecapping parameterization of Alves and Banner[19] but in fact based on different physics than the original formulation of Alves and Banner, as argued by Babanin and Westhuysen [20]. The reason to choose this formulation is based on our operational SWAN implementation validations for the Black Sea [16] and some studies for the Mediterranean coast of France [21-22]. Our conclusions, based on SWAN validation using satellite altimetry data and buoy data (for the French coast) are that the parameterization approach of Westhuysen is giving the lowest RMSE and bias with regards to the wave period (actually significantly lower for the case of the French coast, where is possible to test the period forecast) while for the significant wave height (Hs) the bias and RMSE are comparable with Komen formulation and significantly better than the Janssen formulation. Conclusions about the applicability of the formulation have been presented by Rusu [12],[23] and the conclusions there are in good agreement with our conclusions about the performance of Westhuysen formulation.

Wave model run is for the period 1996-2003, using ERA Interim wind forcing data and ALADIN model data for 2003. The output is compared with an estimation of the mean wave power, based on the measurements at the buoy near Gelendzhik, Russia due to the fact that these are the only available wave measurements for a long enough time period. While these measurements are for the Eastern Black Sea Shelf, this way some comparison with actual measurements can be made. Based on the measurements the mean monthly and annual wave power has been estimated and compared with the SWAN wave model output for the entire period of available measurements. Wave energy potential of three locations in the Bulgarian shelf has been also estimated (well known to be the most energetic places there), but there are significant concerns about the reliability of the results.

**RESULTS AND DISCUSSION**

The location of the sites, for which the estimations will be presented, are shown on fig.1

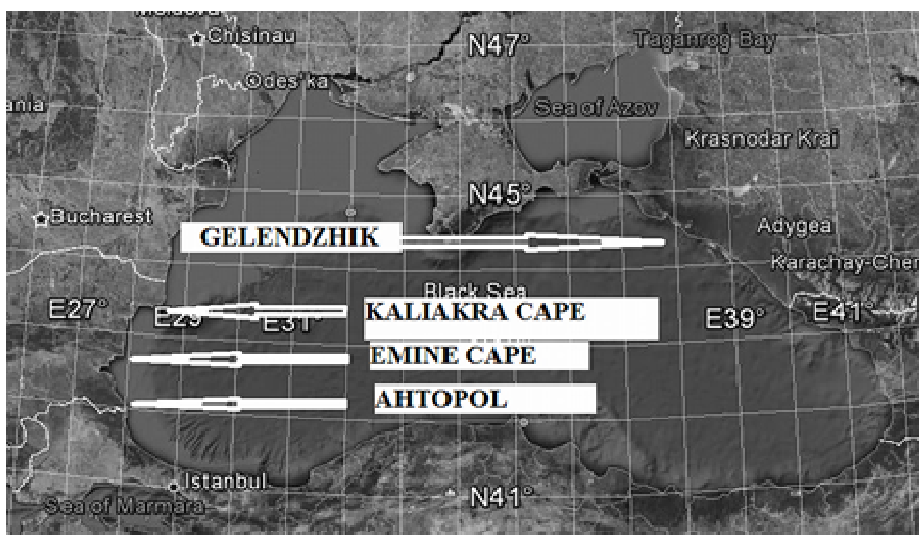

**Fig.1.** Locations of the sites of interest

SWAN model run was carried for the period 1996-2006 and mean monthly wave power has been calculated (excluding some months for which there are gaps in wave measurements). Next the actual monthly mean wave power has been calculated, based on the measured significant wave height and mean period from the buoy near Gelendzhik. On fig.2 the distribution of the monthly means comparing SWAN with the measured mean wave power flux is presented. The mean annual wave power potential estimated using SWAN for the buoy position is 1.72 kW/m while the one obtained from the measurements is 1.82 kW/m but while these results are in a good agreement, from fig.2 is clear that the SWAN simulations (forced by ERA Interim) underestimate the wave energy during the winter months (there is a significant underestimation of the highly energetic conditions) and overestimate the summer months wave energy (the low energy conditions). While for the buoy locations this still leads to an overall realistic estimation of the mean annual value, the issue with the underestimation for the Western shelf can be bigger, due to the fact that it is the most energetic shelf, while the Eastern shelf is the least energetic and so the underestimation can lead to much more significant underestimation of the mean annual wave power potential. On fig.3 the share of the waves with different wave heights to the total wave power is presented.

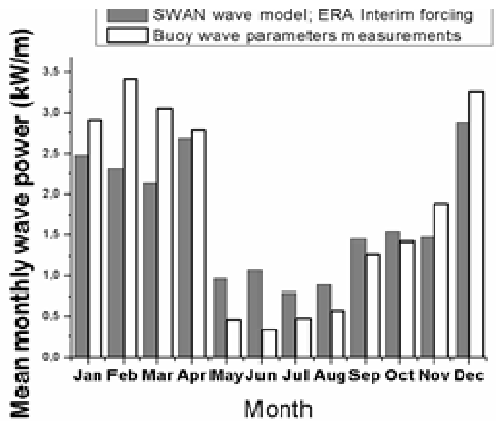

**Fig.2.** Monthly mean wave power for the Gelendzhik buoy location using SWAN output and the measured.

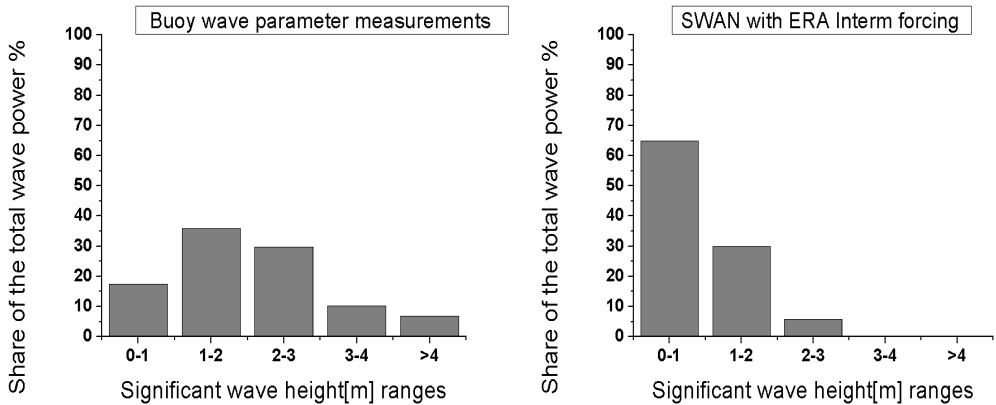

**Fig.3.** Share of the energy of waves with different wave heights to the total wave energy in %

While from the measurements there is a maximum due to the waves with significant wave height which is 1-2m Hs, the model output is with a very different distribution and the share of the waves above 3m Hs is negligible and the reason is the underestimation of the strong winds by ERA Interim and overestimation of the wind speeds during low energetic conditions.

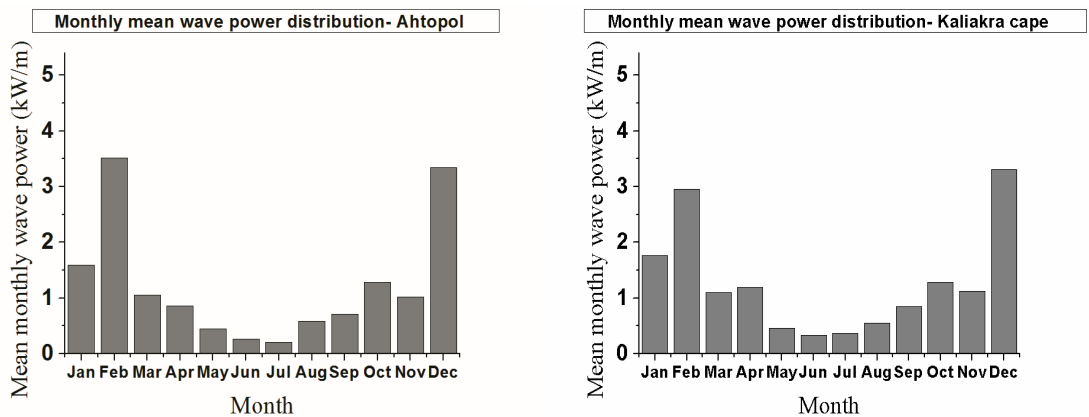

**Fig.4.** Monthly mean wave power distribution for Ahtopol and Kaliakra cape (Bulgarian shelf)

Next the mean monthly and annual wave power potentials for the three Bulgarian sites have been estimated and the results are presented on fig.4. The mean annual wave power potential using ERA Interim forcing is 1.24 kW/m for Ahtopol, 0.99 kW/m for Emine cape and 1.27 kW/m for Kaliakra cape (all output points in the model for these sites are selected at 50m depth in order to have estimation without decrease due to wave transformation). These values are significantly lower than the values that can be deduced if we take into account the wave climate studies of the Bulgarian shelf [24],[25].

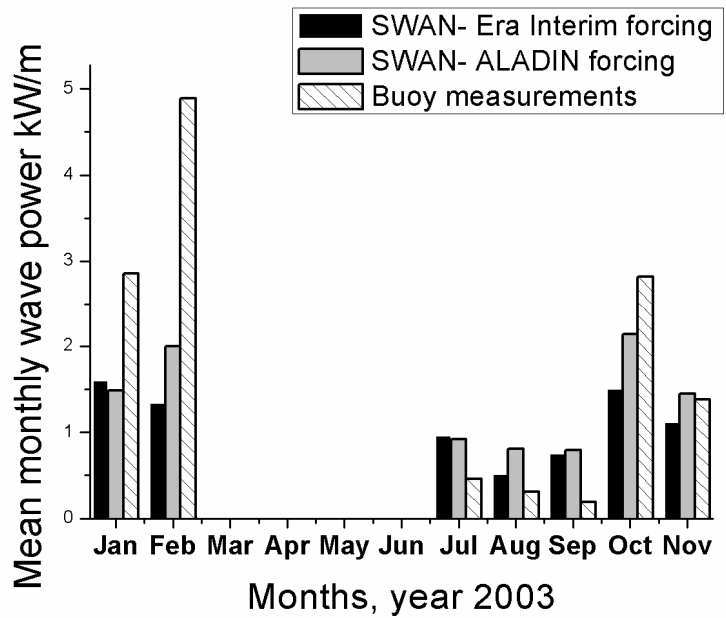

**Fig.5.** Mean monthly wave power potential for 7 months in 2003, comparison of the values calculated using the measurements and SWAN with ALADIN and ERA forcing.

As a next step we ran SWAN for the 7 months in 2003 with available wave measurements, forced by the ALADIN model [26] wind fields with a spatial resolution of 0.25° (now it is 0.125° over the Black Sea). On fig.5 the monthly means are compared. In general SWAN with ALADIN forcing performs slightly better than ERA Interim+SWAN but again we observe a significant underestimation of the modeled wave power for the winter months. For these 7 months the mean wave power for the buoy location calculated from the measurements is 1.84 kW/m, the simulated by SWAN+Era Interim is 1,11 kW/m and by SWAN+ALADIN is 1.34 kW/m.

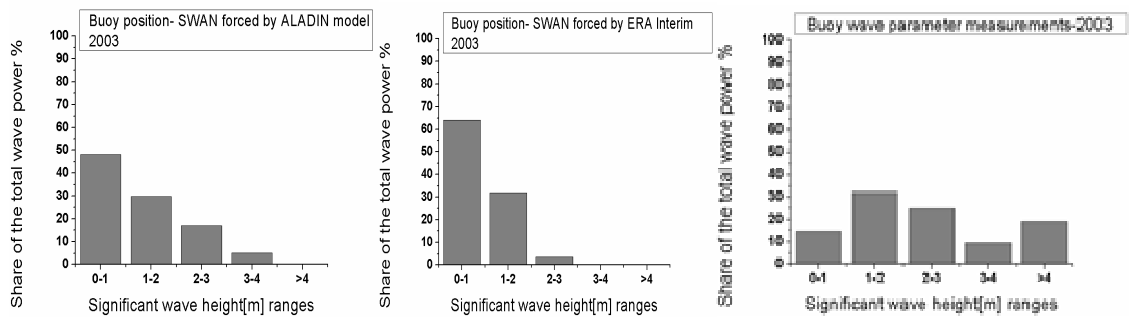

**Fig.6.** Share of the wave power of the waves with different Hs of the total wave energy for the year 2003. Left- SWAN+ALADIN, center- SWAN+Era Interim, right- measured

We will also mention that the bias and RMSE of the monthly means calculated using SWAN+ Era Interim compared with the calculated by the measurements, and while the bias is reasonable -0.10 kW/m, the root mean square error (RMSE) is 1.09 kW/m which is in the order of the value of the estimated wave power potentials themselves.

Based on the presented results, we will discuss the estimations presented in [14],[15]. The most energetic part of the Bulgarian shelf is well known to be the southernmost part of the shelf (confirmed also by the mentioned articles). Our study location there is nearby the town Ahtopol. We estimated using ERA winds a mean annual wave power potential as 1.24 kW/m but consider it as an underestimated. Akpinar et al[14] estimated that it is in the order of 1.5 kW/m (which corresponds well with our estimation, not surprisingly because we use the same wind input). The mean annual Hs for Ahtopol according to Akpinar is about 0.5m which according to our experience as an operational providers of marine forecasts is unrealistic low (moreover it is obvious from the presented validation time series that the implementation of SWAN is missing all the peaks). According to the estimation of Grozdev [24] based on the decades of visual observations of the wave heights at the coastal weather stations of NIMH- BAS (including a synoptic station at Ahtopol) the mean annual Hs is above 1m nearby the shore and the mean annual Hs is obviously more and even if the visual observations are not reliable as the instrumental measurements, still they are usable enough to conclude that the estimations of the mean annual wave power potential by the use of ERA Interim wind fields are underestimated twice or more. Moreover Akpinar et al [14] concluded that the mean annual wave power potential offshore in the deep part of the sea is 3 kW/m which is obviously too low. Aydogan et al [15] estimated the mean annual wave energy power potential of a point that is near Ahtopol to be 6.58 kW/m and the wave power potential of the point close to Kaliakra cape (referred by the authors as Dobrich because it is in the Dobrich region) 4.82 kW/m. Their estimation of the mean wave power potential of a point close to the Romanian Gloria platform is roughly 4 kW/m. It is possible to compare that value with the wave measurements data, provided by Rusu [12]- based on the data in his article, the mean wave power at that location is 2.0-2.5 kW/m (rough estimate based on the presented wave climate data) which means that while Akpinar and Komurcu (and our own estimation) underestimate about twice, there is an overestimation of the wave power potentials for the Western Black Sea shelf. For the point close to Gelendzhik the estimated mean wave power potential in[15] is 2.11 and so the overestimation is probably not more than 20% for the less energetic Eastern shelf, but for the Western it is probably more. The overall opinion of the author of the present study is that the values presented here and in [14] are the lower possible limit while the values in [15] are not the actual, but the upper possible limit of the wave power potential (and not unlikely to be unreachable) of the Black Sea and the truth is in between but probably closer to the estimate presented by Aydogan et al.

**CONCLUSIONS**

The overall conclusion of the presented study is that the problem of the Black Sea wave energy potential estimation up to now has not been solved. Estimations based on ERA reanalysis or possibly other reanalysis (with low spatial resolution) will produce realistic estimates for the Eastern and Southeastern Black Sea shelf, but not for the Western and Southwestern (the most energetic and promising in terms of possible wave energy conversion). The only way to use reanalyzed wind fields are to use statistical or dynamic downscaling of the reanalysis or some attempt to use statistical corrections (but for the later measurements are needed which are not available). The other possible approach is to use not directly the wind fields, but the mean sea level pressure fields, based on the conclusions of Davidan et al [27] that the wind fields reconstructed by the mean reanalyzed pressure fields are more accurate than the wind fields in the reanalysis for the purposes of wave modeling. The next approach is to use a high resolution regional atmospheric model and our future plans include an estimation based on ALADIN data for the recent 5 years (due to the higher spatial and temporal resolution of ALADIN for that period).